\documentclass[doublespace,11pt]{article}
\usepackage{graphicx}
\usepackage{setspace}
\usepackage{subfigure}
\usepackage{epsfig}

\usepackage{url}

\begin{document}
\sloppy
\begin{spacing}{1}

%\begin{titlepage}
%\hspace{0.08in}
%\begin{minipage}{\textwidth}
%\begin{center}
%\vspace*{3cm}
%\begin{tabular}{c c c}
%\hline
% & & \\
% & {\Huge The University of Algarve} & \\
% & & \\
% & {\Huge Informatics Laboratory} & \\
% & & \\
%\hline
%\end{tabular}\\
%\vspace*{2cm}
%{\Large
%UALG-ILAB\\
%Technical Report No. 200701\\
%September, 2007\\
%}
%\vspace*{3cm}
%{\bf A Sudoku Game for People with Motor Impairments}\\

%\addvspace{0.5in}
%{\bf St\'{e}phane Norte}\\%, and {\bf Fernando G. Lobo}\\
%\vspace*{-0.1in}
%\vspace*{4cm}
%Department of Electronics and Informatics Engineering\\
%Faculty of Science and Technology \\
%University of Algarve \\
%Campus de Gambelas\\
%8000-117 Faro, Portugal\\
%URL: {\verb http://www.ilab.ualg.pt }\\
%Phone: (+351) 289-800900\\
%Fax: (+351) +351 289 800 002 \\
%\end{center}
%\end{minipage}
%\end{titlepage}

\title{\bf A Sudoku Game for People with Motor Impairments}
\author{    {\bf St\'{e}phane Norte}\\
            \small UAlg Informatics Lab\\
            \small DEEI-FCT, University of Algarve\\
            \small Campus de Gambelas\\
            \small 8000-117 Faro, Portugal\\
            \small snorte@ualg.pt
%\and
%\footnote{Also a member of IMAR - Centro de Modela\c{c}\~{a}o Ecol\'{o}gica.}\\
%            {\bf Fernando G. Lobo}\\
%            \small UAlg Informatics Lab\\
%            \small DEEI-FCT, University of Algarve\\
%            \small Campus de Gambelas\\
%            \small 8000-117 Faro, Portugal\\
%            \small flobo@ualg.pt
}
\date{}
\maketitle

\begin{abstract}
Computer games are motivating and beneficial in learning different educational skills.  
Most people use their fingers, hands, and arms when using a computer game. However, 
for people with motor disabilities this task can be a barrier. We present a new Sudoku 
game for people whose motion is impaired, called Sudoku 4ALL. With this special 
interface a person can control the game with the voice or with a single switch. 
Our research aims to cautiously search for issues that might be appropriate for computational 
support and to build enabling technologies that increase individuals' functional independence 
in a game environment.

\end{abstract}

\section{Introduction}
Over the years computer games have expanded, increasing attention focus on games 
accessibility. The Game Accessibility Special Interest Group (GA-SIG) of the 
International Game Developers Association encourages researchers to investigate 
computer games to assist disabled people~\cite{GA-SIG}. There are many types of disabilities 
and limiting conditions that affect a pleasant human-computer interaction.
The primary categories encountered in gaming are limitations in vision, hearing, 
mobility, or cognitive issues.  

In this paper, we will focus on the main problem for 
people with motor impairment to access the Sudoku puzzle. We have created 
a system, called Sudoku 4ALL. With this interface we provide an easy way to help 
people to play the popular Sudoku game. 
Unfortunately, many people with motor disabilities present severe limitations  
when interacting with the computer. With a virtual 
environment we can offer the possibility to operate in the real world, partly alleviating 
physical limitations~\cite{Kerr:02}. People with motor impairments can use alternative devices to interact 
with computer games. Some of the major alternative input technologies for people with 
disability may include switch devices, head-mounted pointing devices, speech-recognition systems, 
eye tracking devices and gloves devices. 

There are a variety of reasons to improve the accessibility of games. A person who has a disability 
should have the same access to equal services and leisure as others in society. In the same 
way, why can't we have games that are accessible to those with disabilities? This becomes a 
quality of life issue. 
There is a lot of work in accessibility technology and in building computers more accessible to 
citizens with disabilities and learning difficulties. However, there is a reduced amount of work 
focused on making all games universally accessible to all, regardless of disability.

In the next section, we describe the accessibility in games design analysis. Then, section 3 presents 
Sudoku 4ALL game, development of the game and 
interaction design. Finally, we describe future work in section 4, and provide our conclusions in section 5.  

\section{Accessibility in Games Design Analysis}

Designing interfaces for people without disabilities is already a difficult task. However, with disabled people 
the methodology becomes even more complex. It is importante to identify the abilities and limitations of the users. 
Specially cognitive and perceptual abilities are relevant to design but 
sensor and motor abilities are important as well.
We provide a variety of possible approaches in the creation of accessible game design to supply persons 
with disabilities, such as:

\begin{itemize}

\item Interfaces must provide various features to adapt to different users' requirements in terms of size, colour, 
contrast and number of items displayed on the screen.

\item Improved hardware supports a wide range of input devices, such as, mice, joysticks, switches, trackballs, 
gloves and microphones.

\item Support a variety of output approach, including text, graphics, sound and speech output.

\item Provide extra attention and concentration on the task. Some designs can break concentration with needless 
distractions, such as animations, popup windows, and intrusive sound effects.

\item Provide simple designs which are more practical in that they avoid and minimize failures in modes that have complex 
designs. Simple designs can be quickly understood and thus support instantaneous use, or encourage further exploration.
 
\end{itemize}

In order to design an interface with good quality and accessibility it is essential to understand the capacity and behavior of 
its users. Simplifying an interface is one of the most important principles in interface design~\cite{Nielsen:95,Norman:02}.

\section{The Game: Sudoku 4ALL}
\begin{figure}
\begin{center}
\includegraphics{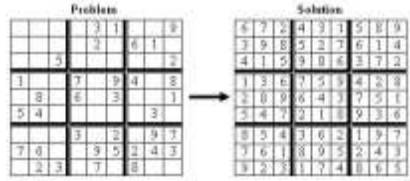}
\end{center}
\caption{A Sudoku puzzle and its solution. The goal is to fill in the blanks with entries 
1-9 so that each digit occurs once in each row, once in each column, and once in each of 
the nine 3x3 squares indicated by bold lines.}
\label{fig:SudokuExample}
\end{figure}

 The Sudoku game is a puzzle of 81 squares written on a 9x9 board where each 
 row and each column contains the digits 1 to 9. 
 The rule of the game is simply to fill in the puzzle board so that the numbers 1 
 through 9 occur exactly once in each row, column, and 3x3 box. The numbers can 
 appear in any order and diagonals are not considered. The initial game board 
 will consist of several numbers that are already placed. Those numbers cannot be 
 changed. The goal is to fill in the empty squares following the simple rule above (see Figure~\ref{fig:SudokuExample}). 
 To solve it, doesn't require any special math skills or calculations. It is a simple and fun game of logic,
 all that's needed is concentration.

The usual 81 cell Sudoku grid is not the only possible board. With any positive integer \emph{a}, 
we can represent an order-\emph{a} Sudoku grid with \emph{a$^{2}$} rows, \emph{a$^{2}$} column and \emph{a$^{2}$} blocks. 
The grid has a total of \emph{a$^{4}$} cells, which are to be completed with numbers in the range from 1 to \emph{a$^{2}$}. 
Difficult Sudoku puzzles can be generalized to \emph{$3^{2} * 3^{2}$} to \emph{$a^{2} * a^{2}$}.

Many people reach a point in the solution process at which they make an intelligent guess about a new fill 
in the grid, and follow the effects of that guess to the solution. Sometimes an 
error occurs in solving the puzzle and forces backtracking search. The use of backtracking can be viewed 
as a logical process. When a person makes a supposition fills in one cell and realizes that some other cell has no correct 
solution, then he has discovered a logical relation between the cells. Most 
people to make a fill in a cell in a Sudoku computer game used a conventional mouse or keyboard, 
but the interaction with people with motor disabilities can be a large problem.
To help people with motor impairment to play the Sudoku puzzle we have created 
a new interface, called Sudoku 4ALL. The Sudoku 4ALL contains special input 
methods to provide important support in using the computer to interact with the game~\cite{Gilligan:06}.

\subsection{Development of the Game}

\begin{figure}
\begin{center}
\includegraphics{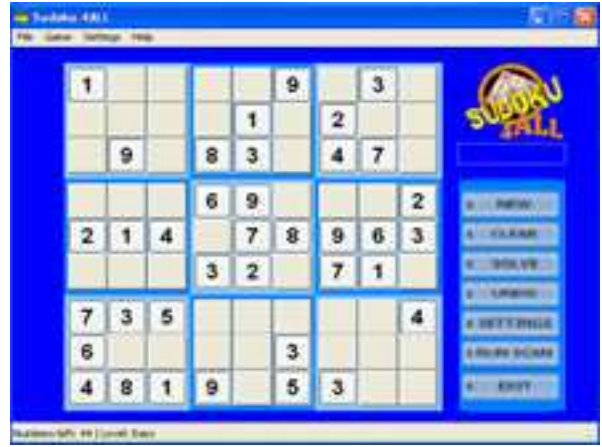}
\end{center}
\caption{The Sudoku 4ALL Interface.}
\label{fig:SudokuInterface}
\end{figure}

The Sudoku 4ALL contains an adjusted environment to support 
different kinds of aids. The interface works with the keyboard, switch control and speech 
recognition. The design of the interface has been carefully developed to provide an easy interaction 
for handicapped people. Figure~\ref{fig:SudokuInterface} shows the Sudoku 4ALL interface. 
The interface contains organized buttons to provide an easy utilization. The goal of the button called \emph{NEW}
is to provide the possibility to generate a new Sudoku game; the \emph{CLEAR} button is to clean the 
entire fill in the cells. The \emph{SOLVE} button provides the solution of the Sudoku grid and the \emph{UNDO} 
button provides backtracking and cleans the last fill in a cell. The \emph{SETTINGS} button allows a user to 
configure several options of the Sudoku. The \emph{RUN SCAN} button is to activate the scanning method. 
Scanning is used to describe the process of moving between and selecting items from a selection set 
using a switch or switches. Items or groups of a selection set are highlighted in turn over time. The 
scanning system allows the users to use a numeric keyboard to provide an easy method to send numbers inside the Sudoku cells.
Finally, the \emph{EXIT} button closes the application.

All the features of the puzzle can be used and configured without an assistant's help. 
The user can configure several options of the Sudoku, such as difficulty level, scanning velocity, 
number of repeat scanning cycles, scanning sound, scanning color, input device (mouse, switch, space 
key, speech recognition), and sudoku size.   
All these options can be controlled with a single switch device or by a standard mouse.

\subsection{Interaction Design}

Sudoku 4ALL is the first Sudoku puzzle game version to utilize switch access devices and speech recognition to help 
people with motor impairment.
Different techniques were used when building an application of this sort.

\subsubsection{Switch Access}

\begin{figure}
\begin{center}
\includegraphics{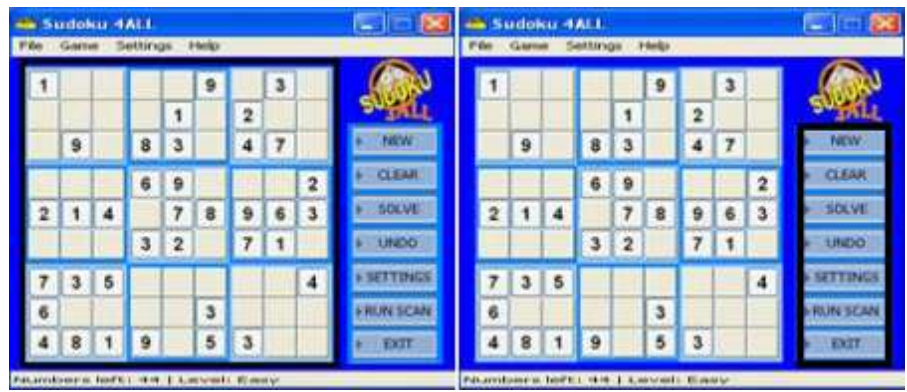}
\end{center}
\caption{An example of the scanning system. There are 2 major groups: the first is to select the Sudoku grid and the
second is used to select the menu options.}
\label{fig:SudokuScanGroups}
\end{figure}

\begin{figure}
\begin{center}
\includegraphics{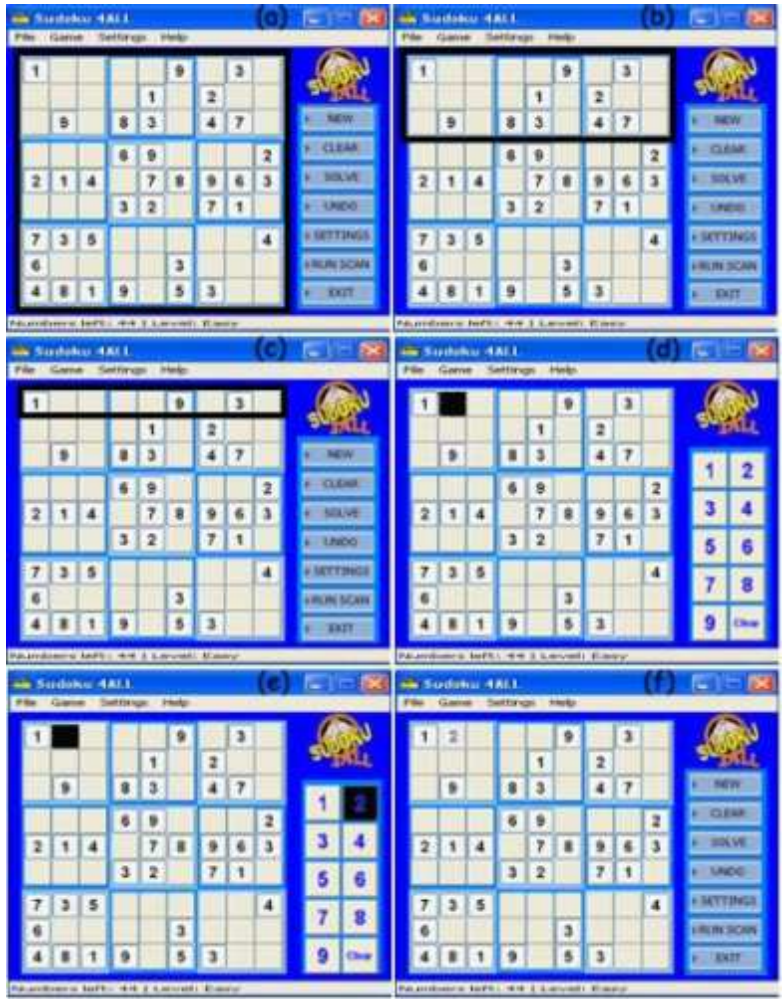}
\end{center}
\caption{a) The first group is focused, b)The first subgroup is selected, c) The first row of the first subgroup is
selected, d) The scanning system becomes focused on the selected items of that row, e) Another scan is activated 
in the numeric keyboard in this phase the next click will be the choice of the desired number, f) Finally, the number 
is sent to the desired cell.}
\label{fig:SudokuAllScan}
\end{figure}

One of the techniques that are usually used to help people with motor impairment to interact with 
computers is the so called scanning method. We used the scanning process in the Sudoku 4ALL to 
allow the user to employ a single key or switch to make choices. The idea is to move from one item 
or group to another after a predefined time. 
There are two principal scanning groups as shown in Figure~\ref{fig:SudokuScanGroups}. The first is used to select the Sudoku 
grid; the second is used to select the menu options. After the user has selected one of the groups, 
the whole selection set is divided into groups of items and these are then individually highlighted (focused). 
The user firsts selects a group, after which the individual items in that same group are scanned. The 
hierarchy of subgroups is scanned until one reaches the level of single items~\cite{snorte:07a}. 
When the item is selected on the Sudoku cell the desired cell is highlighted and waits for the user to make 
the numeric choice. At this stage, another scan is activated in the numeric keyboard in this phase the 
next click will be the choice of the desired number. To use the scanning system, the user needs to use a 
single switch or press the ``Space'' key. Figure~\ref{fig:SudokuAllScan} shows the process related above.

Through the scanning mechanism, the Sudoku puzzle can be controlled by a user with motor impairment. 
The \emph{SETTINGS} option is not an exception. Each feature is controlled by another scanning system. 
The user can configure the options with a single switch or keyboard. At this stage, the scanning proceeds 
at the top of the column from one item to another, in this phase the next click will be the choice of the 
desired option configuration. %Figure~\ref{fig:SudokuSettings} shows an example to configure the scanning colour with 
%only two switch presses.

%\begin{figure}
%\begin{center}
%\includegraphics{SudokuSettings.eps}
%\end{center}
%\caption{In this example, the scanning colour was selected, and now the scanning system becomes focused on the items 
%of that group, with the cursor advancing through each of the colour options one at a time.}
%\label{fig:SudokuSettings}
%\end{figure}

\subsubsection{Voice Access}

\begin{figure}
\begin{center}  
\includegraphics{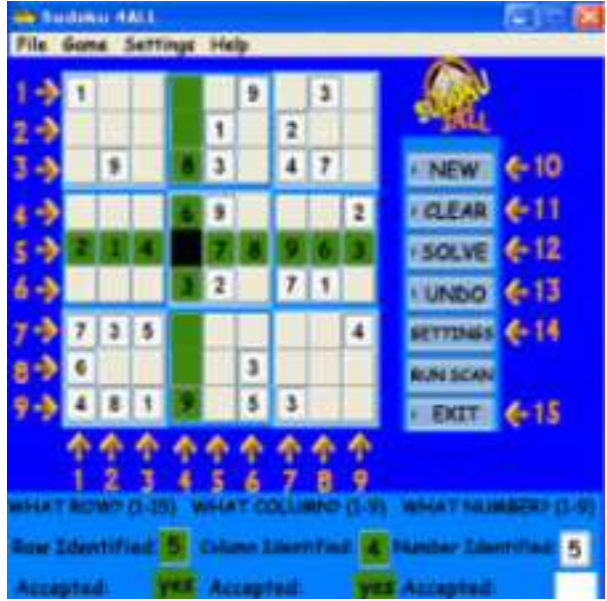}  
\end{center}
\caption{Using the voice the user selects the desired row (1-15) and column (1-9). Finally, the next step 
is to choose the number in the Sudoku grid.}
\label{fig:SudokuVoice}
\end{figure}

The utilization of speech recognition can be used in different areas to help people with motor 
disabilities~\cite{Karimullah:02}. We provide, in the Sudoku 4ALL a speech-to-text system to interact with the Sudoku 
grid and configure several options. To insert numbers inside the grid the users need to select the 
desired row (1-9), if the chosen row is wrong, the user then says ``No''. If it is correct, ``Yes''. If the answer is 
positive the next step is to choose the column (1-9) using the same ideas described before. Finally, the 
user can say a number to put inside the grid. Figure~\ref{fig:SudokuVoice} shows an example of the Sudoku 4ALL controlled by voice.
A user can configure other features in the Sudoku 4ALL using numbers 
between 10 and 15. If the user says ``10'' a new Sudoku game will be generated. If she pronounces the number ``11'' the 
entire fill in the cells will be deleted; the number ``12'' provides the solution to the puzzle and the number ``13'' removes 
the last fill in a cell.  
The number ``14'' supplies the user to configure the difficulty level of the puzzle, the row and column color, Sudoku size and provides 
the configuration of the Microsoft speech recognition training wizard. After choosing the number ``14'' the user 
needs to pronounce other numbers between 1 and 5 to choose the desired options.
To close the game the user needs to pronounce the number ``15''.

\section{Future work}

In the near future, we intend to continue to test the Sudoku 4ALL with more individuals 
with motor impairment including people with cerebral palsy. We would like to continue to improve
the Sudoku 4ALL with new options to help people with visual impairments.
%Our current goal is primarily focused on expanding the use of educational interfaces among people with disabilities.

\section{Conclusions}

This paper presents the Sudoku 4ALL, a system to help individuals with physical disabilities 
to play the Sudoku puzzle. The Sudoku 4ALL is available for free use and can be downloaded 
from \url{http://w3.ualg.pt/~snorte/Sudoku4ALL.htm}. We hope that the Sudoku 4ALL will be useful to many 
people, and encourage researchers to explore new ways to aid users with disabilities.

\section*{Acknowledgments}
This work was sponsored by the Portuguese Foundation for Science and 
Technology (FCT/MCTES) under grant \url{POCI/CED/62497/2004}.

\bibliographystyle{abbrv}
\bibliography{references}

%\vfill\eject

\end{spacing}
\end{document}